\documentstyle[aps,epsfig,multicol,rotate]{revtex}
\def\beginwide{
        \end{multicols} \vspace*{-0.5cm} \noindent
        \rule{3.5in}{.1mm}\rule{.1mm}{5mm} \widetext \medskip }
\def\beginwidetop{
        \end{multicols} \vspace*{-0.5cm} \noindent
        \widetext \medskip }
\def\endwide{
        \hspace*{3.35in}~\rule[-5mm]{.1mm}{5mm}\rule{3.5in}{.1mm}
        \begin{multicols}{2} \vspace*{-1.0cm} \noindent }
\def\endwidebottom{
        \begin{multicols}{2} \vspace*{-1.0cm} \noindent }

\begin{document}
\title{Flux front penetration in disordered superconductors} 

\author{Stefano Zapperi$^{1}$, Andr\'e A. Moreira$^{2}$ 
and Jos\'e S. Andrade Jr.$^{2}$}
\address{$^1$INFM sezione di Roma 1, Dipartimento di Fisica, Universit\`a 
``La Sapienza'', P.le A. Moro 2, 00185 Roma, Italy.
$^2$Departamento de F\'{\i}sica, Universidade Federal do
Cear\'a, 60451-970 Fortaleza, Cear\'a, Brazil.
}

\date{\today}

\maketitle
\draft
\begin{abstract}
We investigate flux front penetration in a disordered type 
II superconductor by molecular dynamics (MD) simulations
of interacting vortices and find scaling laws for
the front position and the density profile. The scaling 
can be understood performing a coarse graining  
of the system and writing a disordered 
non-linear diffusion equation. Integrating numerically the 
equation, we observe a crossover from flat to fractal
front penetration as the system parameters are varied.
The value of the fractal dimension indicates that the 
invasion process is described by gradient percolation.
\end{abstract}


\begin{multicols}{2}

The magnetization properties of type II
superconductors have been studied for many years,
but the interest in this problem has been renewed
with the discovery of high temperature superconductors \cite{Blatter}.
The magnetization process is usually described in terms of the 
Bean model \cite{bean} and its generalizations: flux lines enter
into the sample and,  due to the presence of quenched disorder, 
give rise to a steady flux gradient. 
While the Bean model provides a consistent picture
of average magnetization properties, such as the hysteresis
loop and thermal relaxation effects \cite{kim}, it does not account for 
local fluctuations in time and space. 
It has been recently observed that flux line
dynamics is intermittent, taking place in avalanches 
\cite{avalanches},  and flux fronts are   
not smooth \cite{rough,fractal,fractal2}. 
In particular, it has been shown that the flux 
front crosses over from flat to fractal 
as a function of material parameters 
and applied field \cite{fractal}.

A widely used modeling strategy to describe
the fluctuations around the Bean state consists
in molecular dynamics (MD) simulations of interacting flux
lines, pinned by quenched random impurities 
\cite{Brass,nori1,nori2,nori3,pla,nori4}. With this
approach it has been possible to reproduce flux profiles \cite{nori2}, 
hysteresis \cite{nori2}, avalanches \cite{nori1,pla,nori4} 
and plastic flow \cite{Brass,nori4}. One of
the aims of these studies \cite{nori2} is to establish precise connections 
between the  microscopic models and the macroscopic behavior,
captured for instance by generalized Bean models. A different 
approach treats the problem at mesoscopic scale, describing the
evolution of interacting coarse-grained units \cite{coarse,discrete}, 
supposed to represent the system at an intermediate scale.
While these models give a faithful representation of several
features of the problem, the connection with the underlying
microscopic dynamics remains unexplored.

In this letter, we investigate the invasion of magnetic
flux into a disordered superconductor, a problem that 
has recently been the object of intense experimental research
\cite{rough,fractal,fractal2}. We first analyze the problem
by MD simulations, in analogy with Ref.~\cite{nori2}, 
and obtain scaling laws relating the front position
and the flux profile to the pinning strength. In order to
understand these results, we perform a coarse-graining 
of the equation of motion and obtain a non-linear diffusion
equation for the flux line density.
In the absence of pinning, the equation 
reduces to the model discussed in Refs.~\cite{dorog,gilc}.
This model has been analytically \cite{dorog,gilc}
solved to provide expressions for 
the initial dynamics of the front that are in agreement 
with our MD simulations. We show
that when quenched disorder is included in the diffusion equation, 
the flux front roughens and is eventually pinned. 
Varying the parameters of the model 
(applied field, disorder, interaction strength), 
the fluctuations of the front display 
a crossover from flat to fractal
that is consistent with experimental observations
\cite{rough,fractal,fractal2}. The value of the fractal dimension 
suggests that the strong disorder limit is described by 
percolation. 

In an infinitely long cylinder, flux lines can be modeled 
as a set of interacting particles performing 
an overdamped motion in a random pinning
landscape \cite{Brass,nori1,nori2,nori3,pla,nori4}. 
The equation of motion for each flux line $i$ can be written as
\begin{equation}
\Gamma \vec{v}_i = \sum_j \vec{J}(\vec{r}_i - \vec{r}_j)+
\sum_p \vec{G}[(\vec{R}_p-\vec{r}_i)/l]+\eta(\vec{r}_i,t),
\label{eq:vf}
\end{equation}
where the effective viscosity can be expressed in terms of
material parameters $\Gamma=\Phi_0 H_{c2}/\rho_n c^2$.
Here, $\Phi_0$ is the magnetic quantum flux, $c$ is the speed of light,
$\rho_n$ is the resistivity of the normal phase and $H_{c2}$ is
the upper critical field.  The first term on the right hand side
represents the vortex-vortex interaction and it is given by
$\vec{J}(\vec{r})\equiv[\Phi_0^2/(8\pi\lambda^3)]K_1(|\vec{r}|/\lambda)\hat{r}$
where the function $K_1$ is a Bessel function decaying exponentially
for $|\vec{r}| > \lambda$ and $\lambda$ is the London penetration
length \cite{degennes}.
The second term on the right hand side accounts for the 
interaction between pinning 
centers, modeled as localized traps, and flux lines.
Here, $\vec{G}$ is the force due to a pinning center located at
$\vec{R}_p$, $l$ is the range of the wells (typically $l \ll \lambda$),
and $p = 1, ..., N_p$ ($N_p$ is the total number of pinning centers).
While Refs.~\cite{nori1,nori2,nori3} employed parabolic traps,  
we decided to avoid discontinuities in the force  
and used instead $\vec{G}(\vec{x})=-f_0\vec{x}(|\vec{x}|-1)^2$, 
for $|\vec{x}|<1$ and zero otherwise.  
Finally, we add to the model an uncorrelated thermal noise term $\eta$,
with zero mean and variance $\langle \eta^2\rangle = k_b T/\Gamma$.
Although in the present simulations we restrict ourselves to the case $T=0$
(see Ref.~\cite{MON-00} for the implementation of thermal noise
in MD simulations), we will 
discuss the effect of temperature in the coarse grained
description of the dynamics.

We perform MD simulations based on Eq.~(\ref{eq:vf}) and 
analyze the flux front propagation for different values of the
pinning strength $f_0$.  We use $N=5000$ flux lines
in a system of size $(L_x=800 \lambda , L_y=100\lambda$ \cite{nota_size}), with
$N_p=800000$ Poisson distributed pinning centers of
width $l=\lambda/2$, corresponding to a density of
$n_0= 10/\lambda^2$. The injection of  magnetic flux
into the sample is implemented similarly to Ref.~\cite{nori2},
concentrating at the beginning of the simulation
all the flux lines in a small strip, parallel to the $y$ direction,
of length $L'=10^{-2}\lambda$ and imposing periodic boundary conditions
in both directions. The front position $x_p$ is identified
as the $x$ coordinate of the 
most advanced  particle in the system at different times. 
In Fig.~1 we show that 
$x_p$ grows initially with time as $t^{1/3}$
and eventually  saturates to a value $\xi_p$ 
which increases as the strength
of the pinning centers $f_0$ is decreased.
The data collapse shown in the inset of Fig.~1 indicates that the
front pinning length $\xi_p$ scales as $f_0^{1/2}$.
When the front saturates, we measure the density profiles \cite{nori2}
which can be collapsed using the scaling form 
$\rho(x)=f_0^{-1/2}F(xf_0^{1/2})$ (see Fig.~2).
In addition, we vary the density of pinning
centers, using $n_0 \lambda^2=2.5,5,7.5,10$, and find that
similar scaling collapses hold if $f_0$ is replaced 
by $f_0 \sqrt{n_0}$.

In order to understand these results, we perform a coarse graining 
of Eq.~(\ref{eq:vf}), starting from the Fokker-Plank equation 
for the probability distribution
of the flux line coordinates $P(\vec{r}_1,....,\vec{r}_N,t)$
\begin{equation}
\Gamma\frac{\partial P}{\partial t} =
\sum_i \vec{\nabla}_i( -\vec{f}_i P +k_B T\vec{\nabla}_i P),
\label{eq:fp}
\end{equation}
where $\vec{f}_i$ is the force on the particle $i$ given by 
Eq.~(\ref{eq:vf}). Next, we introduce the single particle density
$\rho(\vec{r},t)\equiv\langle\sum_i\delta^2(\vec{r}-\vec{r}_i)\rangle$, 
where the average is done over the distribution 
$P(\vec{r}_1,....,\vec{r}_N,t)$. The evolution of $\rho$ can
be directly obtained from Eq.~(\ref{eq:fp}) and is given by
\begin{eqnarray}
\Gamma\frac{\partial \rho}{\partial t} & = &
-\vec{\nabla}{\Huge (}\int d^2r' \vec{J}(\vec{r}-\vec{r}\;')
\rho^{(2)}(\vec{r},\vec{r}\;',t) \nonumber\\ &  &
-\sum_p \vec{G}[(\vec{R}_p-\vec{r})/l]\rho(\vec{r},t){\Huge )}
+k_BT \nabla^2\rho,
\end{eqnarray}
where $\rho^{(2)}(\vec{r},\vec{r}\;',t)$ is the two-point density,
whose evolution depend on the three-point density and so on.
The simplest truncation scheme involves 
the approximation $\rho^{(2)}(\vec{r},\vec{r}\;',t)
\simeq \rho(\vec{r},t)\rho(\vec{r}\;',t)$. We then 
coarse grain the equation considering length scales larger
than $\lambda$. This can be done expanding $\vec{J}$
in Fourier space, keeping only the lowest order term
in $\vec{q}$, and retransforming back in real space.
The result reads
\begin{equation}
\int d^2r' \vec{J}(\vec{r}-\vec{r}\;')\rho(\vec{r}\;',t)\simeq 
-a\vec{\nabla} \rho(\vec{r},t),
\label{eq:a}
\end{equation}
where $a\equiv \int d^2r \vec{r}\cdot\vec{J}(\vec{r})/2=\Phi_0^2/4$.

The coarse graining of the disorder term is more subtle.
A straightforward elimination of short wavelength modes
would give rise, as in the previous case,  
to a random force $\vec{F}_c(\vec{r})= -g\vec{\nabla}n$,
where $n$ is the coarse grained version of the 
microscopic density of pinning centers 
$\hat{n}(\vec{r})\equiv \sum_p\delta^2(\vec{r}-\vec{R}_p)$
and $g \propto f_0$ . This method can not
be applied for short-range attractive pinning forces as the one
we are investigating. In this case, 
short wavelength modes yield a macroscopic contribution
to pinning that can not be neglected. 
Consider for instance the flow between two coarse grained regions:
short-range microscopic pinning forces give rise to a macroscopic
force that should always oppose the motion, while the random force
derived above could in principle point in the direction of 
the flow. In other words, $F_c(\vec{r})$ should be considered as
a {\em friction} force \cite{nota_bp}
whose direction is always opposed to the driving force $\vec{F}_d$
(in our case $\vec{F}_d=a \vec{\nabla}\rho$) and whose absolute value
is given by $|g\vec{\nabla}n|$ for $|\vec{F}_d| >|g\vec{\nabla}n|$
and to $|\vec{F}_d|$ otherwise \cite{friction}. 

Collecting all the terms, we finally obtain a disordered
non-linear diffusion equation for the density of flux lines 
\begin{equation}
\Gamma\frac{\partial \rho}{\partial t}=
\vec{\nabla}(a\rho\vec{\nabla}\rho-\rho\vec{F}_c)
+k_BT \nabla^2\rho.
\label{eq:fin}
\end{equation}
The boundary conditions representing our MD 
simulations correspond to a constant number 
of flux lines $ML_y$ injected into the system,
so that the total density is conserved $\int dxdy \rho(x,y,t)= M L_y$.
With these boundary conditions, the $T=0$ behavior 
depends on a single effective coupling constant $g_0\equiv g\sqrt{n_0}/(aM)$
as can be shown rescaling Eq.~(\ref{eq:fin}) 
and the conservation law as
\begin{equation}
\rho=\tilde{\rho}/M~~~t=\tilde{t}/(\Gamma aM^2) ~~~n=\tilde{n} \sqrt{n_0}.
\end{equation}

In the limit $g_0=0$ and $T=0$, Eq.~(\ref{eq:fin}) has 
been solved by Bryskin and Dorogotsev
\cite{dorog} using similarity methods and the solution reads 
$\rho(x,y,t)=t^{-1/3} h(x/t^{1/3})$,
where $h(u)=(1-u^2)/6$  for $u<1$ and vanishes for $u\geq 1$ 
\cite{dorog,gilc,nota_prof}. Notice that we recover here the $t^{1/3}$
behavior observed in MD at early times for the front position
\cite{nota_bc}. When disorder is present ($g_0>0$), we expect 
the front to deform and eventually encounter a 
strong pinning region where it stops. 
The scaling of the front position $\xi_p$ with $f_0$ observed 
in MD simulations can be explained, noticing 
that the front will be pinned when the force due
to the density gradient is equal to the friction 
force of the order of $f_0\sqrt{n_0}$. The gradient force
can be estimated as $a\nabla \rho \sim \rho_0/\xi_p$, where 
the density at the boundary is given by $\rho_0 \sim M/\xi_p$.
Thus we obtain that $\xi_p \sim (Ma/f_0\sqrt{n_0})^{1/2}=g_0^{-1/2}$ 
in agreement with MD simulations. 

In order to confirm the validity of these considerations,
we discretize  Eq.~(\ref{eq:fin}) 
on a tilted square lattice and 
integrate the equations numerically using a
finite volume technique with an upwind scheme 
to avoid numerical instabilities. 
At the beginning of the simulation,
all flux lines are concentrated at the boundary of the
sample (i.e. $x=0$) and in each site of the lattice 
we define an uncorrelated Gaussian pinning center density
$n$, with mean $n_0$ and variance $\sqrt{n_0}$, thus
defining a local random friction force. 
As flux lines enter into  the sample, 
we identify the flux front using a burning algorithm
and compute its average position $x_p$.
In the initial stage, $x_p$  grows as 
$t^{1/3}$ up to a crossover length scaling as
$\xi_p \sim g_0^{-1/2}$, in agreement with MD simulations
(see Fig.~3). In addition, we measure the density profiles and
find that they rescale with $g_0$ in the same way as in  
MD simulations.

The numerical integration of the diffusion equation allows  
for a direct analysis of the fluctuations in the front as
a function of different internal parameters. 
Measuring the width $W$ of the fronts as
a function of time for different values of $g_0$, we find
that in the initial stage $W$ grows as a power law $t^\beta$
where $\beta \simeq 0.35$ until it saturates to a value that
decreases with $g_0$. Thus the front crosses over from flat
to fractal as it enters into the material. 
In principle we can control the strength of the fluctuations 
and the associated characteristic length $\xi^*$
by tuning $g_0$, which directly reflects experimentally 
measurable parameters. 

In order to compare the model with experiments, we have
to implement appropriate boundary conditions. In 
Refs.~\cite{rough,fractal} the external field was ramped
at constant rate, which corresponds to a constant increase
of the boundary density: $\rho(0,y,t)=ht$ \cite{gilc}.
Integrating the equation with this boundary condition,
we observe that for strong disorder the flux front roughens
displaying substantial overhangs (see the inset of Fig.~4 \cite{nota_rough}).
Here the box counting method is applied to estimate the 
fractal dimension of the front: we divide the lattice 
in boxes of size $b$ and count the number of occupied
boxes $N(b)$, which decays as $b^{-D_f}$,
where $D_f$ is the fractal dimension. Fig. 4 shows
that for $b<\xi^*$ the data are well fitted
by a power law with exponent $D_f=4/3$, 
which coincides with the fractal dimension
of the perimeter of the percolation cluster. 
For $b>\xi^*$ the fractal dimension crosses 
over to $D_f=1$, corresponding to a flat
front. The observed behavior is reminiscent of gradient percolation
\cite{gradper}, a model in which the concentration
of occupied sites $p$ decreases with the distance 
from the sample boundary. The invaded area is compact, but
its perimeter is fractal (with $D_f=4/3$) up to a length scale
$\xi^*$ which depends on the concentration gradient.

In experiments on flux penetration in Tl$_2$Ba$_2$CuO$_{6+x}$ thin films 
with controlled anisotropy, $D_f$ was  
found to change continuously from $D_{f}\simeq 1.5$ to $D_{f}\simeq 1$ 
as the degree of anisotropy and the 
applied field was varied \cite{fractal}. It would be 
interesting to check if the observed continuous variation
of the fractal dimension can be explained by a crossover from 
a percolation process at short length scales to a flat invasion
at large length scales, with a crossover length  depending
for instance on the degree of anisotropy.

In conclusion, we have analyzed flux front penetration
in disordered type II superconductors, establishing connections
between microscopic flux line dynamics and a mesoscopic non-linear
diffusion equation. This equation allows to interpret
the results of MD simulations and 
provides a useful framework to analyze experiments.

This work has been supported by CNPq and FUNCAP. We thank 
A. Baldassarri, K. Behnia, M. C. Miguel, S. Sarti,
A. Scala and A. Vespignani for useful discussions.
S. Z. wishes to thank for hospitality the physics department 
of UFC where this work has been initiated.

\newpage
\begin{figure}  
\epsfig{file=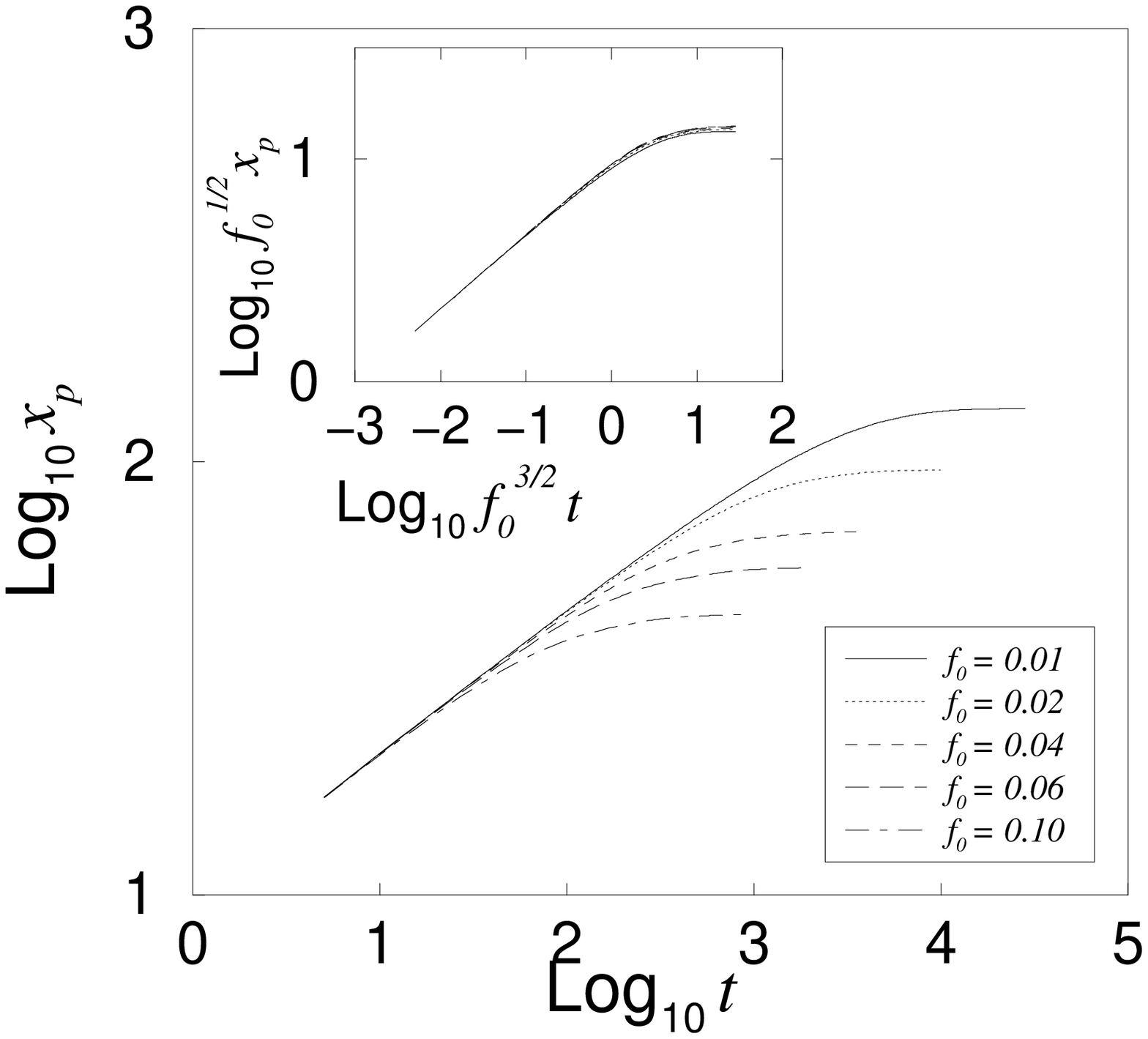,width=8cm} 
\caption{The average position of the front, obtained from MD
simulations injecting a constant
magnetic flux from the boundary, is plotted as a function of time.
The curve increases as $t^{1/3}$ and saturates at long times
to a value depending on  $f_0$.
In the inset, we show by data collapse 
that the pinning length scales as $f_0^{-1/2}$} 
\end{figure}

\begin{figure}  
\epsfig{file=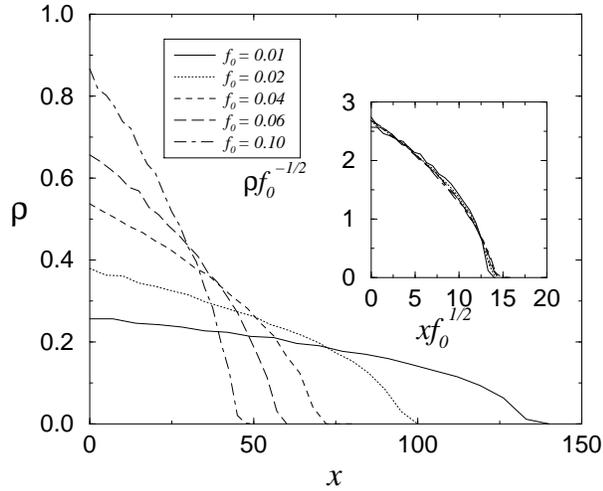,width=8cm} 
\caption{The pinned density profiles measured in MD simulations. In the inset
we show the data collapse.} 
\end{figure}

\begin{figure}  
\epsfig{file=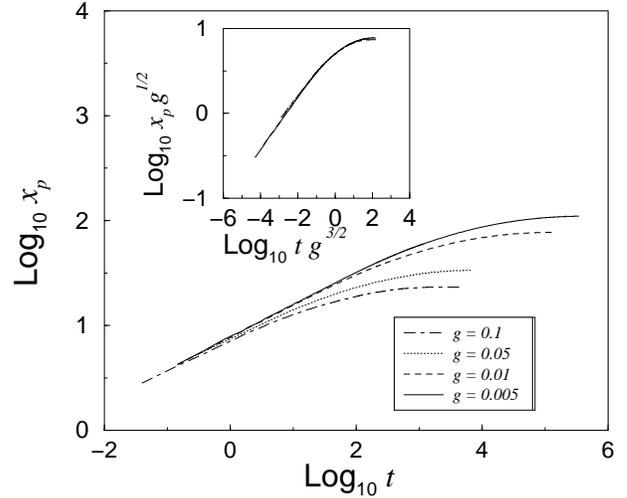,width=8cm} 
\caption{The front position measured simulating the 
continuum equation for different values of $g$.
In the inset we show the data collapse.} 
\end{figure}

\begin{figure}  
\epsfig{file=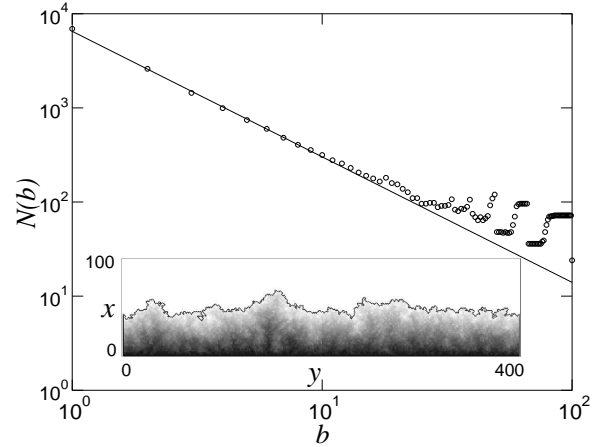,width=8cm,angle=-90} 
\caption{In the inset we show a density plot
obtained simulating the continuum equation with $g=5$, $a=6.28$, $n_0=8$, 
and concentration at the boundary increasing at rate $h=0.01$.
In the main figure we report the box counting plot of the front, averaged
over twelve realizations of the disorder. The front has fractal dimension
$D_f=4/3$ (solid line) and crosses over to $D_f=1$ at large length
scales.} 
\end{figure}

\end{multicols}

\begin{thebibliography}{99}
\bibitem{nota_size}
	The results do not depend on $L_y$ if the fluxline densisty
	is held constant.
\bibitem{nota_bp}
	The lattice model described in Ref.~\cite{discrete}
	implements a random force rather than a friction force,
	which is used instead in Ref.~\cite{coarse}. 
	We have checked	that a random force produces
	a different scaling for the front position
	and the profiles than MD simulations.
\bibitem{nota_prof} This functional form only holds 
	in the initial regime, when the effect of
	disorder is weak.
\bibitem{nota_bc}
	In the continuum equation, 
	the scaling exponent depends on the boundary 
	conditions \protect\cite{dorog,gilc}. We have checked 
	that this feature is also true for MD simulations.
\bibitem{nota_rough}
	Ref.~\cite{rough} observed self-affine scaling
	for the flux front. While we can not exclude that
	for some parameter range self-affine scaling could
	apply, this is not the case for the present data.
\end{thebibliography}
\end{document}